%% file: main.tex
\documentclass[11pt]{article}
\pdfoutput=1 
\usepackage{./jheppub}

\usepackage[T1]{fontenc} 

\usepackage{graphicx}
\usepackage{amsfonts,amsmath,amssymb}
\usepackage{verbatim}
\usepackage{array}
\usepackage{mathrsfs}
\usepackage{mathtools}
\usepackage{yfonts}
\usepackage{dsfont}
\usepackage{bbm}
\usepackage{colonequals}
\usepackage{amscd}
\usepackage{slashed}
\usepackage{extarrows}

\usepackage{float}
\usepackage{afterpage}

\usepackage{wrapfig}

\usepackage{suffix,mathtools,cancel,bbm}

\usepackage{multirow}

\usepackage{enumerate}

	\usepackage{tikz-cd}

\input{./auxx}

\usepackage{colortbl}
\definecolor{mygray}{gray}{0.93}

\usepackage{sectsty}
\allsectionsfont{\boldmath}

\usepackage{comment}

\title{\boldmath $U(1)$ $R$-Symmetry Topological Operators from Branes in Holography}

\author[a]{Thomas Waddleton} 
\affiliation[a]{William H.~Miller III Department of Physics and Astronomy, Johns Hopkins University, Baltimore, MD 21218, USA}

\abstract{We study $U(1)$ symmetry generators from wrapped branes in Type IIB string theory and M-theory. Specifically, we construct topological defects for the $R$-symmetry in 4d $\cN=1$ and 3d $\cN=2$ quantum field theories holographically dual to Sasaki-Einstein compactifications. These are studied using a universal consistent truncation to the gravity multiplet in the bulk theory. We furthermore check these constructions by identifying the objects charged under them, and understanding their origin in the bulk theory.}

\usepackage{etoolbox}
\appto\appendix{\addtocontents{toc}{\protect\setcounter{tocdepth}{1}}}

\appto\listoffigures{\addtocontents{lof}{\protect\setcounter{tocdepth}{1}}}
\appto\listoftables{\addtocontents{lot}{\protect\setcounter{tocdepth}{1}}}

\begin{document} 

\setcounter{tocdepth}{2}

\maketitle
\flushbottom


\input{./sec_introduction}


\input{./sec_branes_as_sym_ops}


\input{./sec_general_SE5}


\input{./sec_general_SE7}


\input{./sec_discussion}


\section*{Acknowledgments}

I am grateful to Mufaro Chitoto, Jonathan Heckman, Patrick Jefferson, and Justin Kulp for insightful conversations and correspondences. I would also like to thank Federico Bonetti, Enoch Leung, and Konstantinos Roumpedakis for comments on an earlier draft. I am especially grateful to Ibou Bah for the original discussions leading to this analysis. TW is supported in part by NSF grant PHY-2112699, and also in part by the Simons Collaboration on Global Categorical Symmetries.


\appendix



\bibliographystyle{./ytphys}
\bibliography{./refs}

\end{document}

%% file: auxx.tex
\newcommand{\bn}{\begin{enumerate}}
\newcommand{\en}{\end{enumerate}}

\def\bZ{\mathbb{Z}}

\def\d{\delta}

\def\det{{\rm det}}

\def \d{\partial}

\newcommand{\beq}{\begin{equation}}
\newcommand{\eeq}{\end{equation}}

\newcommand\nn{\nonumber}

\newcommand{\cA}{\mathcal{A}}

\newcommand{\cD}{\mathcal{D}}
\newcommand{\cE}{\mathcal{E}}

\newcommand{\cG}{\mathcal{G}}

\newcommand{\cI}{\mathcal{I}}

\newcommand{\cM}{\mathcal{M}}
\newcommand{\cN}{\mathcal{N}}

\newcommand{\cR}{\mathcal{R}}

\newcommand{\cU}{\mathcal{U}}
\newcommand{\cV}{\mathcal{V}}

\numberwithin{equation}{section}

\def\bea{\begin{eqnarray}}
\def\eea{\end{eqnarray}}

\DeclarePairedDelimiterX\MeijerM[3]{\lparen}{\rparen}%
{\begin{smallmatrix}#1 \\ #2\end{smallmatrix}\delimsize\vert\,#3}

\newcommand\MeijerG[8][]{%
  G^{\,#2,#3}_{#4,#5}\MeijerM[#1]{#6}{#7}{#8}}

\WithSuffix\newcommand\MeijerG*[7]{%
  G^{\,#1,#2}_{#3,#4}\MeijerM*{#5}{#6}{#7}}

\def\bR{\mathbb{R}}
\def\bZ{\mathbb{Z}}

\def\cN{\mathcal{N}}

\def \beg#1{\begin{#1}} 

\def \bea{\beg{eqnarray}}
\def \eea{\end{eqnarray}}
\def \ee{\end{equation}}

\def \restr#1#2{{\left.\kern-\nulldelimiterspace#1\vphantom{\big|}\right|_{#2}}}

\def \nn{\nonumber}

\def \d{\partial}

\def \sfA{\mathsf{A}}

\def \sfF{\mathsf{F}}

\def \bQ{\mathbb{Q}}
\def \bR{\mathbb{R}}

\def \bZ{\mathbb{Z}}



%% file: sec_introduction.tex

\section{Introduction}\label{sec:introduction}

In the last decade or so, the importance of symmetries in physics has exploded in many new directions. Formalized in \cite{Gaiotto:2014kfa}, the modern approach to describing global symmetries in QFTs is through the language and manipulation of {\it topological} operators: extended submanifolds of spacetime supporting local degrees of freedom in a way that depends only on the topological data of the submanifold. Through this language the notion of symmetry has been generalized greatly to include higher-form symmetries described by $p$-form gauge fields, higher-group symmetries describing more intricate structure between two (or more) symmetry groups \cite{Cordova:2018cvg,Benini:2018reh}, and most generally to non-invertible and {\it categorical} symmetries \cite{Kaidi:2021xfk,Roumpedakis:2022aik}.

A driving tool in the study of generalized global symmetries in a $d$-dimensional QFT is a $(d+1)$-dimensional gauge theory that has the lower-dimensional theory living on its boundary. By analyzing operators defined in the bulk theory along with admissible boundary conditions, much can be learned about the boundary QFT without needing specific details about the theory's local data. The $(d+1)$-dimensional theory cares only about {\it global} data of boundary. For this reason the bulk theory is often referred to as a Symmetry Topological Field Theory (SymTFT) \cite{Burbano:2021loy,Freed:2012bs,Gaiotto:2020iye,Apruzzi:2021nmk,Freed:2022qnc} elsewhere in the literature.

This method of using a higher-dimensional theory to describe a QFT lends itself immediately to {\it holographic} constructions in string theory and M-theory. Indeed, via the now famous AdS/CFT correspondence, physicists have been studying QFTs defined on $d$-dimensional spacetimes using a theory in one dimension higher for decades \cite{Maldacena:1997re}. The SymTFT specifically has been computed for various QFTs directly from holographic constructions \cite{Apruzzi:2021nmk}, building naturally off previous work on anomaly inflow in string and M-theory \cite{Harvey:1998bx,Freed:1998tg,Bah:2018gwc,Bah:2019jts,Bah:2018jrv,Bah:2019rgq}. What's more, in these constructions we are already familiar with the idea of wrapped branes in the bulk geometry describing operators in the boundary theory \cite{Witten:1998xy,Aharony:1998qu}. This idea has been refined more in recent years, including a non-trivial identification that has been proposed between topological operators in the boundary theory and wrapped branes that have had their dynamics decoupled in a suitable limit \cite{Apruzzi:2022rei}. We will discuss this limit in more detail in Section \ref{sec:branes_SymOps} below. While the majority of the results in the literature have focused on examples of discrete symmetries in the boundary theory \cite{Apruzzi:2021nmk,GarciaEtxebarria:2022vzq,Apruzzi:2022rei,vanBeest:2022fss,Apruzzi:2023uma,Bah:2023ymy,Heckman:2022muc,Heckman:2022xgu,Dierigl:2023jdp}, few studies have been performed for continuous symmetries \cite{Cvetic:2023plv,Garcia-Valdecasas:2023mis,Bergman:2024aly}. In the present paper we will further develop the study of continuous symmetry operators from wrapped branes.

Specifically we study two families of examples: Type IIB in an $AdS_5\times SE_5$ background and M-theory in an $AdS_4\times SE_7$ background, where $SE_{2n+1}$ denotes an internal $(2n+1)$-dimensional Sasaki-Einstein manifold \cite{Gibbons:2002th,Morrison:1998cs,Acharya:1998db,Sparks:2010sn}. The defining characteristic of a general Sasaki-Einstein $(2n+1)$-manifold is that its cone is Calabi-Yau. That is, our two examples will focus on branes probing the tip of the cone geometry
\begin{equation}
    ds^2(CY_{2n+2}) = dr^2 + r^2 ds^2(SE_{2n+1}).
\end{equation}
A very important fact for the present paper is that any $SE_{2n+1}$ can be presented as a circle fibration over a K\"ahler-Einstein base $KE_{2n}$. This allows us to write a local expression for the base metric as
\begin{equation}
    ds^2(SE_{2n+1}) = (\eta + \sfA_1)^2 + ds^2(KE_{2n}),
\end{equation}
where $\eta$ is a 1-form dual to the $U(1)$ Killing vector, known as the Reeb vector, and $\sfA_1$ is a gauge field associated to this fibration. In the above expression, the 1-form $\eta$ is not closed but instead satisfies $d\eta = 2J$, where $J$ is the K\"ahler 2-form defined on the base $KE_{2n}$.

In the case of Type IIB string theory, we are concerned with the theory of D3-branes probing a general cone $CY_3$ over an $SE_5$ base. There is an additional piece of data for an $SE_5$ coming from the complex structure of the K\"ahler-Einstein base: a (2,0)-form that we will denote $\Omega$. This form is not closed but satisfies
\beq
d\Omega = 3i\eta\wedge\Omega.
\eeq
The forms $\eta,J,$ and $\Omega$ specify an $SU(2)$ structure group on the $SE_5$. When placed in this background, Type IIB string theory is known to be holographically dual to a 4d $\cN=1$ SCFT, the exact content of which is determined by the chosen $SE_5$. Initial examples of these constructions included the well-known cases where the internal space was described by the round metric\footnote{Strictly speaking this is not dual to an $\cN=1$ SCFT but rather $\cN=4$ SYM, but the round metric on $S^5$ is still Sasaki-Einstein.} on $S^5$ \cite{Maldacena:1997re} and the homogeneous metric on $T^{1,1}$ \cite{Klebanov:1998hh}, before later expanding greatly to include the countably-infinite families of solutions described by $Y^{p,q}$ \cite{Gauntlett:2004yd,Martelli:2004wu,Benvenuti:2004dy,Buchel:2006gb} and $L^{a,b,c}$ \cite{Cvetic:2005ft,Cvetic:2005vk} metrics. Independent of this choice there will always be a global $U(1)$ symmetry in the 4d SCFT coming from the $R$-symmetry, which is understood holographically as coming from the $U(1)$ isometry present in any $SE_5$ geometry. The global symmetry we will be focused on constructing the topological defect for is this ``universal'' $R$-symmetry.

The second example that we will discuss is in M-theory, but is geometrically very similar to the IIB construction above. Our background geometry will be described by a stack of M2-branes probing the tip of a Calabi-Yau 4-fold cone $CY_4$ over an $SE_7$ base. There is again additional data coming from the complex structure of the base $KE_6$: a $(3,0)$-form $\Omega$ describing the complex structure. This form satisfies
\beq
d\Omega = 4i\eta\wedge\Omega.
\eeq
This endows the $SE_7$ manifold with an $SU(3)$ structure coming from $\eta, J$ and $\Omega$. In this background, M-theory is holographically dual to a 3d $\cN=2$ SCFT in a way that depends on the internal $SE_7$. Similar to the $SE_5$ case, the number of known metrics for $SE_7$ manifolds for many years was quite small \cite{Freund:1980xh,Duff:1986hr}, before later expanding to include an infinite family of metrics known as $Y^{p,k}$ \cite{Gauntlett:2004hh,Martelli:2008rt,Kim:2010ja}.\footnote{We specify that the Freund-Rubin geometries of the form $AdS_4\times S^7$ as well as the later generalization to the ABJM geometries $AdS_4\times S^7/\bZ_k$ \cite{Aharony:2008ug}, while not strictly dual to $\cN=2$ SCFTs, are still of Sasaki-Einstein type.} In all of these cases however, there will always be a global $U(1)$ $R$-symmetry in the dual 3d $\cN=2$ SCFT. Its holographic origin is understood as the same as the previous case: the isometry of the circle fiber in the internal $SE_7$ manifold. We will construct a topological operator for this ``universal'' $R$-symmetry similar to the Type IIB case.

This paper is organized as follows. In Section \ref{sec:branes_SymOps}, we review and discuss our main tool for the analysis: Page charges derived from constraints in the bulk string theory. We also discuss the process where D$p$-branes in Type IIB string theory and M-branes in M-theory can be used to describe topological operators in a boundary QFT. Specifically, we discuss the decoupling of dynamics on these branes, leaving behind only topological data. The following two sections present examples for constructing $U(1)$ $R$-symmetry generators from wrapped branes probing the Sasaki-Einstein backgrounds described above. In Section \ref{sec:general_SE5} we construct a D3-brane wrapping the $S^1 \subset SE_5$ fiber that reproduces the $R$-symmetry generator in the boundary field theory. In Section \ref{sec:general_SE7} we consider the analogous symmetry in an M-theory construction, find that its corresponding operator is described by an M2-brane wrapping the $S^1\subset SE_7$ fiber, and comment on the relation to the Type IIB construction. In Section \ref{sec:discussion} we discuss the holographic origin of these symmetry operators as singleton modes of the bulk theory, and conclude by proposing some interesting future research directions. 

%% file: sec_branes_as_sym_ops.tex

\section{Branes to Symmetry Operators}\label{sec:branes_SymOps}

In this section we review the general program of describing topological defects in holographic QFTs using dynamical branes. We first introduce our main tool in identifying the symmetry generators in the boundary theory, Page charges derived from bulk Gauss Law constraints, and then discuss the process by which {\it dynamical} branes can be identified with the {\it topological} symmetry generators.

\subsection{Page Charges and Worldvolume Actions}

In a general QFT, one is interested in describing global $p$-form symmetries through the presence of topological operators in the theory supported on codimension-$(p+1)$ manifolds \cite{Gaiotto:2014kfa}. Such topological operators can be constructed in many ways, with the most explicit being as a holonomy of a closed $(d-p-1)$-form. This is most familiar in the case of continuous $p$-form symmetries which, by Noether's theorem, have associated $(p+1)$-form currents $J_{p+1}$ satisfying a conservation equation $d\star J_{p+1} = 0$. One can construct an operator
\begin{equation}
    U_\alpha(M^{d-p-1}) = \exp\left(2\pi i\alpha\int_{M^{d-p-1}}\star J_{p+1}\right),
\end{equation}
where $\alpha$ parameterizes the symmetry group. This can be shown to be topological using an application of Stokes's Theorem as well as the current conservation equation.

A more general approach to constructing topological operators in a QFT comes from quantities known as Page charges \cite{Page:1983mke,Marolf:2000cb}. These Page charges $P_i$ are closed forms that are quantized but not necessarily gauge-invariant. That the $P_i$ are closed immediately causes their holonomies to only depend on the topological class of the manifold they are defined over, and that they are quantized causes these holonomies to be single-valued. In the case where a Page charge $P_i$ is not gauge-invariant, one can supplement the holonomy with additional ``statistical fields'': degrees of freedom localized to the operator's support that allow one to write the operator in a gauge-invariant way \cite{Choi:2022jqy,Choi:2022fgx}. 

The use of Page charges to describe gauge symmetry transformations is not a recent idea and has its roots in holography \cite{Moore:2004jv,Belov:2004ht}. Indeed, one can understand the relation directly in holographic settings where our $d$-dimensional QFT of interest is on the boundary of a $(d+1)$-dimensional bulk gauge theory as follows. In the bulk theory, one can perform Hamiltonian quantization along a codimension-1 slice and derive Gauss Law equations from the non-dynamical modes of the gauge fields. These manifest as constraints $\cG_i = 0$ that need to be imposed on the boundary, or along any timeslice chosen for quantization. The statement of gauge invariance of the boundary theory can be written as the invariance under the action of the operators
\begin{equation}
    \exp\left(2\pi i\int_{\partial\cM^{d+1}}\lambda\wedge \cG_i\right),
\end{equation}
where $\cM^{d+1}$ is the bulk spacetime, $\cM^{d}$ denotes our codimension-1 slice used for quantization, and $\lambda$ is the gauge parameter of appropriate form degree. If $\cG_i$ is an exact form, we write $\cG_i = dP_i$ and consider singular gauge transformations such that the above exponential can be written as a holonomy,
\begin{equation}
    \exp\left(2\pi i\alpha \int_{M^p}P_i\big|_{\cM^{d}}\right).
\end{equation}
By ``singular gauge transformation'' we are considering $d\lambda = \alpha\,\delta(M^p)$ for $\alpha$ a constant depending on the symmetry group. We then see that holonomies of Page charges naturally arise from considering gauge invariance of the bulk theory defined with our QFT of interest on the boundary. More recently, this idea has been connected to the construction of topological operators in such setups \cite{Apruzzi:2022rei}. The holonomies of Page charges are identified with operators in the boundary theory; that these objects are topological can be seen from the Gauss Law constraints $\cG_i = dP_i = 0$. Explicit examples are the focus of later sections.

With the previous motivation of finding Page charges to construct topological defects, we now discuss the origin of the $(d+1)$-dimensional bulk gauge theory. For the holographic constructions we are interested in, the bulk can be obtained from the 11d (12d) anomaly polynomial associated to string theory (M-theory) \cite{Bah:2018gwc,Bah:2019jts,Bah:2020jas}. One reduces the anomaly polynomial over the compact internal dimensions to obtain a theory only in the external $AdS_{d+1}$ space. It is this final theory that defines the effective bulk theory used to get the necessary Page charges \cite{Apruzzi:2022rei,Bah:2023ymy}.

Once one has obtained the Page charges and thus topological operators in a holographic QFT, a natural question to ask is what is their origin in the dual string theory? The proposed answer is that the topological defects in a QFT are dual to dynamical branes in the bulk theory that have been pushed all the way to the conformal boundary. The explicit details of this identification differ depending on the bulk theory and will be discussed in what follows, but we first make some general arguments. 

For a given brane in the bulk theory, it will generally have a non-trivial quantum field theory living on its worldvolume. This theory will have a sector encoding dynamics living on the worldvolume as well as a sector that is purely topological and depends only on the brane's shape through the pullback of bulk degrees of freedom. The sector dictating dynamics is characterized by the brane tension, which we denote $T_p$ for a $(p+1)$-dimensional brane. As the brane is pushed to the conformal boundary in a holographic construction, the tension will scale with the radius, $T_p \sim r^p$, causing the brane to become infinitely heavy. This scaling of the tension causes the dynamics to ``freeze out'' and leave behind a purely topological worldvolume action living on the brane \cite{Apruzzi:2022rei,Bah:2023ymy}. In many known cases so far, the remaining action can be identified as a Page charge in the dual field theory, thus allowing one to identify the bulk brane configuration with a boundary topological operator.

\subsection{D$p$-branes in Type II String Theory}

In Type II string theory, the dynamical branes are in part described by D$p$-branes. The different sectors of a D$p$-brane worldvolume theory are captured by two parts: the Dirac-Born-Infeld (DBI) action and the Wess-Zumino (WZ) action. The former encodes the dynamics of worldvolume scalars and gauge fields, while the latter describes the topological couplings of bulk RR and NSNS gauge fields coming from anomaly inflow. As was heuristically described above, the DBI action has an overall coupling given by the brane's tension,
\begin{equation}
    S_{\rm DBI}^{{\rm D}p} = T_p\int d^{p+1}\sigma\sqrt{\det\left(G_{ab} + B_{ab} + F_{ab}\right)},
\end{equation}
where $\sigma^a$ denotes the worldvolume coordinates, the fields $G_{ab}$ and $B_{ab}$ are the pullback of bulk fields to the brane worldvolume, $F_{ab}$ is the curvature for the worldvolume Chan-Paton bundle, and we are implicitly using units where $2\pi\alpha' = 1$. As a D$p$-brane is pushed to the conformal boundary in a holographic setting, this tension will diverge and cause the DBI action to decouple from the worldvolume theory. 

After decoupling the dynamics, the only sector that remains is that described by the WZ action (see, e.g. \cite{Polchinski:1995mt,Polchinski:1996fm}).
\begin{equation}\label{eq:IIB_WZ}
    S_{\rm WZ}^{{\rm D}p} = \mu_p\int_{\cM^{p+1}}I_{p+1}^{{\rm D}p} := \mu_p\int_{\cM^{p+1}}\sum_{q\leq p}\left.C_{q+1}\wedge {\rm ch}_B(\cE)\wedge\sqrt{\frac{\hat{\cA}(\cR_T)}{\hat{\cA}(\cR_N)}}\right|_{(p+1)-{\rm form}}
\end{equation}
Here, $\mu_p$ is the RR charge for a single D$p$-brane, $C_{q+1}$ are pullbacks of the bulk odd- (even-) degree RR fields in Type IIA (IIB) string theory, and ${\rm ch}_B(\cE)$ is the Chern character for the Chan-Paton bundle twisted by the NSNS 2-form over the worldvolume $\cM^{p+1}$. The quantities $\hat{\cA}(\cR_T)$ and $\hat{\cA}(\cR_N)$ respectively denote the $A$-roof genus of the tangent and normal bundles of $\cM^{p+1}$; we will omit these gravitational coupling in the analysis that follows for simplicity.

We now take a moment to recall that the field strengths of the RR fields are obtained from the potentials via a differential twisted by $H_3$, the field strength for the NSNS 2-form $B_2$.\footnote{Indeed, this twisting is one of the main supporting arguments that the RR fields, and by extension the charges for D$p$-branes, are classified by twisted K-theory \cite{Minasian:1997mm,Witten:1998cd}.} Namely, the RR field strengths are defined locally as
\begin{equation}
    G_{q+2} = dC_{q+1} + H_3\wedge C_{q-1},
\end{equation}
and satisfy a non-trivial Bianchi identity
\begin{equation}
    dG_{q+2} = -H_3\wedge G_{q}.
\end{equation}
These Bianchi identities will be intimately related to the Page charges we derive in the later examples. Due to the electromagnetic duality enjoyed by Type II string theory relating $G_q$ and $G_{10-q}$, the Bianchi relations above also encapsulate the equations of motion obtained from the low-energy 10d SUGRA actions, and so we do not have additional constraints to consider.

\subsection{M-branes in M-theory}

In M-theory, there are fewer known dynamical objects to work with as there is only a single 3-form RR potential $C_3$; there are electrically-charged M2-branes and magnetically-charged M5-branes. For a single M2-brane, the different sectors are again captured by two parts: a 3d Nambu-Goto (NG) action and the WZ action coupling to $C_3$ \cite{BERGSHOEFF198775,Bergshoeff:1987qx,simon_brane_2012}. The NG action has an overall scaling from the tension of the M2-brane, as seen below.
\begin{equation}
    S_{\rm NG}^{{\rm M}2} = T_{{\rm M}2}\int d^3\sigma \sqrt{\det(G_{ab})}
\end{equation}
Here $\sigma^a$ are the worldvolume coordinates of the M2-brane and $G_{ab}$ is the pullback of the bulk metric to the brane. Similar to the D$p$-branes above, as we push an M2-brane to the conformal boundary, the tension will diverge and freeze the dynamics of the worldvolume theory. The analogous action for M5-branes, however, is difficult to describe quantitatively due to the presence of a self-dual 2-form field that lives on the worldvolume.\footnote{For some progress in this direction, see \cite{Bandos:1997ui,Berman:2007bv,Mathai:2014aya,Giotopoulos:2024sit}.} Fortunately we do not need an explicit expression, as we are focused on a limit where the dynamics freeze out and we are left with only a topological action. To this end all that is relevant is an overall scaling by the tension of the M5-brane, which is indeed present.

Once the NG action decouples, the remaining topological sectors of the M2- and M5-brane worldvolume actions are described by relatively simple WZ actions.
\begin{equation}\label{eq:Mtheory_WZ}
    S^{{\rm M}2}_{\rm WZ} = \mu_{{\rm M}2}\int_{\cM^3}C_3,\qquad S_{\rm WZ}^{{\rm M}5} = \mu_{{\rm M}5}\int_{\cM^6}C_6 + \frac{1}{2}H_3\wedge C_3
\end{equation}
In the above equations $\mu_{{\rm M}2}$ and $\mu_{{\rm M}5}$ respectively denote the RR charge of a single M2- and M5-brane, $C_3$ is the pullback of the bulk RR potential to the brane worldvolume, $C_6$ is the electromagnetic dual gauge field of $C_3$, and $H_3$ is the field strength of the self-dual worldvolume 2-form.

In contrast to Type II string theory, the RR field strength of M-theory is obtained by a standard differential. Locally, we may write
\begin{equation}
    G_4 = dC_3,
\end{equation}
so that $G_4$ has a standard Bianchi identity $dG_4 = 0$. However, there is a non-trivial relation coming from the equations of motion for $C_3$. From the low energy 11d SUGRA action, one can find that $G_4$ must satisfy
\begin{equation}
    d\star G_4 = -\frac{1}{2}G_4\wedge G_4,
\end{equation}
which, along with the Bianchi relation, will give us the Page charges we are looking for in later sections.

%% file: sec_general_SE5.tex

\section{General Sasaki-Einstein in IIB}\label{sec:general_SE5}

Our first example is that of the $R$-symmetry present in $\cN=1$ SCFTs constructed from Type IIB string theory on an $AdS_5\times SE_5$ background. As described in Section \ref{sec:introduction}, this $R$-symmetry is universal in the sense that its presence can be understood as coming from the $U(1)$ isometry in the internal $SE_5$ when viewed as an $S^1$ fibration over a $KE_4$ base. We start with a general consistent truncation of Type IIB supergravity and truncate further to a multiplet containing only the gauge field associated to this isometry. From there we can identify the $U(1)$ operator in the boundary field theory as well as the wrapped brane configuration that produces it.

\subsection{Type IIB KK Reduction}

To construct the operator generating the $U(1)$ $R$-symmetry in a universal way, we start with an arbitrary $AdS_5\times SE_5$ background in Type IIB string theory. Our starting Kaluza-Klein ansatz for the metric of Type IIB supergravity is given by \cite{Liu:2010sa,Gauntlett:2010vu,Cassani:2010uw,Bah:2010cu}
\beq
ds_{10}^2 = e^{-\frac{2}{3}(4U+V)}ds^2(\cM^5) + e^{2U}ds^2(KE_4) + e^{2V}\left(\eta+\sfA_1\right)^2,
\eeq
where $\cM^5$ is an arbitrary external 5-manifold, $U$ and $V$ are scalar fields parameterizing the breathing and squashing modes of the $KE_4$, and $\sfA_1$ is defined over the external spacetime. The internal geometry only describes a Sasaki-Einstein 5-manifold if the breathing and squashing modes are not present, so we focus on the case $U=V=0$.

The RR fields are expanded most generally in this ansatz as
\begin{align}
G_5 =\;& 4e^{Z_0}{\rm vol}(\cM^5) + \star_5 K_2\wedge J + K_1\wedge J\wedge J\nn\\
&+\left[2e^{Z_0} J\wedge J - 2\star_5 K_1 + K_2\wedge J\right]\wedge\left(\eta+ \sfA_1\right)\nn\\
&+\left[\star_5 L_2\wedge \Omega + L_2\wedge\Omega\wedge\left(\eta+\sfA_1\right)+{\rm c.c}\right],\nn\\
G_3 =\;& g_3 + g_2\wedge\left(\eta+\sfA_1\right) + g_1\wedge J + g_0 J\wedge\left(\eta+\sfA_1\right)\nn\\
&+\left[N_1\wedge \Omega + N_0\Omega\wedge \left(\eta+\sfA_1\right)+{\rm c.c}\right]\nn\\
H_3 =\;& h_3 + h_2\wedge\left(\eta+\sfA_1\right) + h_1\wedge J + h_0 J\wedge\left(\eta_1+\sfA_1\right)\nn\\
&+\left[M_1\wedge \Omega + M_0\Omega\wedge \left(\eta+\sfA_1\right)+{\rm c.c}\right]\nn\\
G_1 =\;& da_0\nn\\
\Phi =\;& \phi_0,
\end{align}
where $K_i, g_i, h_i, Z_0, a_0,$ and $\phi_0$ are real-valued while $M_i, N_i, $ and $L_2$ are complex-valued. The subscripts in all cases denote the form degree of the field.

From here, we wish to analyze only the {\it universal} data in an $\cM^5\times SE_5$ background. That is, we wish to truncate the external fields to only consider the 5d $\cN=2$ gravity multiplet, containing the metric as well as the $R$-symmetry gauge field $\sfA_1$. This amounts to setting\footnote{The authors of \cite{Gauntlett:2010vu} set $e^{Z_0} = 0$ for the truncation rather than simply $Z_0=0$. However, this causes the equations of motion to no longer be consistent, so we are taking the truncation listed at present.}
\beq
h_i = g_i = M_i = N_i = K_1 = L_2 = Z_0 = a_0 = \phi_0 = 0,\qquad K_2 = -d\sfA_1.
\eeq
In this truncation, all RR field strengths vanish except for the self-dual $G_5$, which takes the form
\begin{align}\label{eq:IIB_5form}
G_5 &= 4{\rm vol}(\cM^5) + 2J\wedge J\wedge (\eta+\sfA_1) - \sfF_2\wedge J\wedge (\eta+\sfA_1) - \star_5\sfF_2\wedge J\nn\\
&= 4{\rm vol}(\cM^5) + 2J\wedge J\wedge (\eta+\sfA_1) -2\sfA_1\wedge J\wedge J +(\star_5 \sfF_2+\sfA_1\sfF_2)\wedge J\nn\\
&\quad + d\Big((\star_5\sfF_2 + \sfA_1\sfF_2)\wedge(\eta+\sfA_1) -\sfA_1\wedge J\wedge(\eta+\sfA_1) + \sfF_2\wedge J + \sfA_1\wedge\star_5\sfF_2\Big).
\end{align}
The final term above gives a globally-defined expression for $C_4$, while the others give a cohomologically non-trivial flux coming from D3-branes in the background geometry.

As a special case, we can recover the expression for a $T^{1,1}$ background by taking 
\beq
J = -\frac{1}{2}(V_1+V_2),\qquad \eta = d\psi + \frac{1}{4\pi}\sum_{i=1}^2\cos\theta_i d\phi_i.
\eeq
A part of $T^{1,1}$ backgrounds, coming from the definite topology, is that we may also turn on external gauge fields corresponding to nontrivial cohomology classes of the internal manifold along the usual lines of a K\"unneth decomposition \cite{Cassani:2010na}. In general, non-trivial cycles in the internal geometry can result in additional gauge fields to consider. These gauge fields will be in different multiplets than the gravity multiplet we are considering here however, so we may safely omit them from this analysis.

Now that we have the field content of our Type IIB background, we wish to find the symmetry operators present in the dual field theory. Specifically, we are interested in the symmetry operator for the $U(1)$ $R$-symmetry. As described in Section \ref{sec:branes_SymOps}, in order to find the relevant Page charge in our 4d theory, we should construct a 5d theory in the $\cM^5$ bulk. We start from the 11d anomaly polynomial for Type IIB string theory derived in \cite{Bah:2020jas}.
\begin{equation}
    \cI_{11} = \frac{1}{2}\cG_5\wedge d\cG_5 + \cG_5\wedge H_3\wedge G_3
\end{equation}
Here, $\cG_5$ is defined via the relation $G_5 = (1+\star_{10})\cG_5$ describing the self-duality of the RR 5-form. In the case at hand, only the first term is nonzero. We next reduce over the internal $SE_5$ geometry using the expression for $G_5$ in \eqref{eq:IIB_5form} to find the 5d bulk theory\footnote{Throughout we will take $\cM^5$ to be a Spin manifold. On such manifolds, the integral $\int\sfA_1\sfF_2\sfF_2$ is valued in $6\bZ$ \cite{Witten:1996qb}.}
\begin{equation}\label{eq:IIB_bulk}
    S_{5d} = \int_{\cM^5} \frac{1}{2}\sfF_2\wedge\star_5\sfF_2 - \frac{1}{2}\sfA_1\wedge\sfF_2\wedge\sfF_2,
\end{equation}
where $\sfF_2$ is the field strength of the gauge field $\sfA_1$.\footnote{Note that this 5d bulk theory is not strictly topological due to the Maxwell term. This is similar to the Symmetry Theory proposed in \cite{Apruzzi:2024htg}.} We may now perform a timeslice quantization of this theory to isolate the non-dynamical modes, from which we find the Gauss Law constraint, and thus our desired Page charge.
\begin{equation}
    \cG_{\sfA_1} = d\star_5\sfF_2 + \sfF_2^2 = 0\quad \Rightarrow\quad P_{\sfA_1} = \star_5\sfF_2 + \sfA_1\sfF_2
\end{equation}
It is worth noting that we may also have obtained this constraint directly from the Type IIB Bianchi identities, which in the current example is simply $dG_5 = 0$. We may then obtain a symmetry generator by considering the holonomy of the Gauss Law constraint
\begin{equation}\label{eq:IIB_operator}
    \cU_\alpha(M^3) = \exp\left(2\pi i \int_{\cM^4}\lambda_0\,\cG_{\sfA_1}\right) = \exp\left(2\pi i\alpha\int_{M^3} P_{\sfA_1}\big|_{\cM^4}\right),
\end{equation}
where $\lambda_0$ is a parameter for a small gauge transformation. Here we have considered a singular transformation such that $d\lambda_0 = \alpha\, \delta(M^3)$ for some submanifold $M^3\subset \cM^4$ and some constant $\alpha$ parameterizing the volume of the space transverse to $M^3$. Due to the quantization of $P_{\sfA_1}$, we see that $\alpha\in[0,1)$.

In the above expression, the reader may notice an apparent issue: the Page charge $\left.P_{\sfA_1}\right|_{\cM^4}$ is not a gauge invariant quantity along arbitrary submanifolds $\cM^4 \subset \cM^5$, and thus $\cU_\alpha(M^3)$ itself is not generally gauge-invariant \cite{Damia:2022bcd}. We can alleviate this in two ways; the first is through the introduction of statistical fields living on the defect worldvolume, such as in \cite{Choi:2022jqy,Damia:2022bcd}. That is, in the case $\alpha = \frac{1}{N}$ with $N$ an integer, we can consider coupling our operator \eqref{eq:IIB_operator} to a $U(1)_N$ theory via
\beq\label{eq:FQH_operator}
    \cU_{\alpha=1/N}(M^3) = \int \cD a\, \exp\left(2\pi i \int_{M^3}\frac{1}{N}\star_5\sfF_2 + \frac{N}{2}ada + a\sfF_2\right),
\eeq
where $a$ is a statistical gauge field defined only on $M^3$. One can perform the path integral over $a$ to verify the equivalence of expressions \eqref{eq:IIB_operator} and \eqref{eq:FQH_operator} for $\alpha = \frac{1}{N}$. More generally, for $\alpha = \frac{p}{N}$, we can stack the operator with a TFT $\cA^{N,p}[\sfF_2]$ coupled to the bulk through the field strength $\sfF_2$.\footnote{The theory $\cA^{N,p}[\sfF_2]$ is the minimal 3d TFT that has a $\bZ_N$ global symmetry, which can be defined using inflow from the bulk theory \cite{Hsin:2018vcg,Choi:2022jqy}.} This allows us to define a gauge-invariant operator for $\cU_\alpha(M^3)$ along any slice of our bulk $\cM^5$, but at the cost of restricting $\alpha$ to only take rational values. The operator \eqref{eq:FQH_operator} was discussed \cite{Damia:2022bcd}, where it was shown to generate a $\bQ/\bZ$ 1-form symmetry in the 5d bulk theory.

The second way to fix the lack of gauge invariance in the Page charge is by considering the operator to be defined on the boundary $\cM^4 = \d\cM^5$ rather than an arbitrary codimension-1 submanifold of the bulk. Here, we supplement the definition of $\cU_\alpha(M^3)$ with suitable boundary conditions for the bulk gauge fields. In particular, as we are presently focused on a boundary theory where $\sfA_1$ is a background gauge field, we may pick Dirichlet boundary conditions so that $\delta\sfA_1\big|_{\partial\cM^5} = 0$. This causes \eqref{eq:IIB_operator} to be manifestly gauge-invariant in the boundary theory, as there are no gauge transformations to consider. Since the boundary conditions are independent of the value of $\alpha$, we can realize the full $U(1)^{(0)}$ symmetry on the boundary in this way. 

It is worth noting that the two methods above yield the same operator when considered on $\d\cM^5$; the expression \eqref{eq:FQH_operator}, when supplemented with Dirichlet boundary conditions for $\sfA_1$, reduces to \eqref{eq:IIB_operator}. As was discussed in \cite{Damia:2022bcd}, this presents an interesting puzzle
between the two different descriptions as follows. We may take any operator constructed using statistical fields and push it to $\d\cM^5$ to recover an operator generating a $\bQ/\bZ^{(0)}$ subgroup of the boundary $U(1)^{(0)}$ symmetry. However, we clearly cannot take any boundary $U(1)^{(0)}$ operator and pull it back into the bulk as there are generators with irrational values of $\alpha$ that
cannot be described through stacking with some $\cA^{N,p}[\sfF_2]$. The operators corresponding to irrational $\alpha$ are ``stuck'' to the boundary theory and cannot be consistently pulled into the bulk from the above analysis. It is curious that the full $U(1)$ does not seem to be realized in the bulk, but we do not comment further on this in the present paper.

\subsection{D3-brane on the $S^1$ Fiber}

Now that we obtained an expression for the $U(1)$ $R$-symmetry operator present in a general 4d $\cN=1$ SCFT dual to Type IIB string theory on an $\cM^5\times SE_5$ background, we wish to identify a probe D$p$-brane configuration in the bulk that reproduces this operator when pushed to the conformal boundary. As guiding hints, we are looking for a brane configuration that preserves the $U(1)$ isometry of the circle fibration and upon reduction yields a codimension-2 topological surface in the external spacetime.\footnote{As the symmetry operator is codimension-1 in the 4d QFT, it must be codimension-2 in the 5d bulk as it is pushed {\it parallel} to the conformal boundary.}

The probe configuration we are ultimately interested in is a D3-brane that is wrapping the $S^1$ fiber of the internal $SE_5$. To find our operator, we reduce the Wess-Zumino action \eqref{eq:IIB_WZ} over any internal directions after decoupling the dynamics encoded in the DBI action. Explicitly, the WZ action for a D3-brane wrapping the $S^1$ fiber is reduced as follows.
\beq
\cI_3 = \int_{S^1}C_4 + C_2\wedge(B_2 + F_2) + \frac{1}{2}C_0(B_2+F_2)^2 = (\star_5 \sfF_2 + \sfA_1\sfF_2)
\eeq
Here we have used the expression for $C_4$ derived above, and have made use of the fact that $C_2$, $B_2$, and $C_0$ are zero when truncated to the universal multiplet. The resulting operator is given by the holonomy of $\cI_3$, written
\beq\label{eq:IIB_brane_sym}
\cU_\alpha(M^3) = \exp\left(2\pi i \alpha\int_{M^3}\star_5 \sfF_2 + \sfA_1\sfF_2\right).
\eeq
The constant $\alpha$ appearing in this expression should be understood as follows. As we have seen previously, the integrand of \eqref{eq:IIB_brane_sym} yields the Page charge $P_{\sfA_1}$ and thus should be understood as an element of cohomology $H^3(\cM^5;\bZ)$. In contrast, the external manifold $M^3$ that we are wrapping our D3-brane on is generally a 3-cycle in $H_3(\cM^5;\bR)$. In order for the integral \eqref{eq:IIB_brane_sym} to make sense as a pairing, we must consider a representative of $M^3$ that is valued in integer homology $H_3(\cM^5;\bZ)$. The constant $\alpha$ is simply the factor that arises when considering the projection of $M^3$ to its representative in integer homology. 

As discussed in the previous section, \eqref{eq:IIB_brane_sym} is not a well-defined operator in the bulk $\cM^5$ due to the improper quantization of the Chern-Simons term in its expression. For rational values of $\alpha$, we may instead write the equivalent expression
\beq
    \cU_{\alpha = p/N}(M^3) = \int\cD a\,\exp\left(2\pi i\int \frac{p}{N}\star_5\sfF_2 + \cA^{N,p}[\sfF_2]\right),
\eeq
where $a$ is a gauge field localized to the D3-brane worldvolume. The interpretation of this expression is that there must be {\it induced brane charges} on the D3-brane wrapping $M^3\times S^1$ \cite{Bah:2023ymy}. Specifically, we can understand the coupling to $\sfF_2$ as the presence of induced charge, localized to 1-cycles of $M^3$ dual to the gauge field $a$, originating from D3-branes wrapping the submanifold dual to $J$ in the $KE_4$ base. This gives a physical understanding for the gauge field $a$ as arising from inflow between the D3-brane worldvolumes. 

We may also directly consider the expression \eqref{eq:IIB_brane_sym}, but only once we have pushed the brane to the conformal boundary $\d\cM^5$ and imposed the necessary Dirichlet boundary conditions on $\sfA_1$. Doing this decouples the induced D3-brane charge from the worldvolume, and thus gives a well-defined operator on the boundary for any value of $\alpha \in [0,1)$.

\subsection{Charged Operator from Wrapped D3-brane}

We have above constructed a topological defect in the 4d boundary QFT using a wrapped D3-brane, but have not yet specified which operator is charged under it. To this end, let us consider another D3-brane, this time wrapping the $S^1$ fiber and the two-dimensional manifold dual to the K\"ahler 2-form $J$. We will denote this internal manifold as $N^3\subset SE_5$. In the external spacetime, we extend this D3-brane along the radial direction $\bR_{+}$ of $\cM^5$ so that its worldvolume in $\cM^5$ links with that of the D3-brane discussed above. Near the conformal boundary, we can reduce the WZ action over the internal directions to find, up to a choice of orientation,
\beq
\cI_1 = \int_{N^3}C_4 + C_2\wedge(B_2 + F_2) + \frac{1}{2}C_0(B_2+F_2)^2 = \sfA_1,
\eeq
so that the D3-brane is described by a line operator supporting the bulk gauge field $\sfA_1$,
\beq
\cV_1(\gamma^1) = \exp\left(2\pi i\int_{\gamma^1}\sfA_1\right),
\eeq
where $\gamma^1$ is along the bulk radial direction. We can generalize this to an operator of general integer charge $\cV_n(\gamma^1)$ by simply considering a stack of $n$ D3-branes wrapping the internal space $N^3$. As discussed in \cite{Bah:2023ymy}, any non-Abelian modes in a stack of D$p$-branes decouples as we push the stack to the conformal boundary, and we are left with the center-of-mass modes describing the resulting defect.

As mentioned above, we equip $\sfA_1$ with Dirichlet boundary conditions on the conformal boundary, so that this operator would na\"ively vanish in the boundary QFT. However, we can consider these conditions to be
\beq
\left.\sfA_1\right|_{\partial \cM^5} = d\varphi_0 = 0
\eeq
for $\varphi_0$ a flat $U(1)$-valued scalar in the boundary. Thus, we find on the boundary a local operator $\overline{\cV}_n(x\in \partial\gamma^1)$ from this wrapped D3-brane. This operator is {\it not} topological as the D3-brane is along the radial direction in the $\cM^5$ bulk rather than parallel to the conformal boundary.

We now wish to show that the local operator $\overline{\cV}_n(x)$ is charged under the topological operator $\cU_\alpha(M^3)$ defined previously. To do this, we investigate the braiding of the two D3-branes in the bulk $\cM^5$. This can be done by considering again the timeslice quantization of our bulk action \eqref{eq:IIB_bulk}. Using an analysis similar to that in \cite{Apruzzi:2022rei,Bah:2023ymy}, one finds that the conjugate momentum to the bulk field $\sfA_1$ is exactly
\beq
\frac{\Pi_{\sfA_1}}{2\pi} = \left(\star_5\sfF_2 + \sfA_1\sfF_2\right),
\eeq
so that the arguments of $\cU_\alpha(M^3)$ and $\cV_n(\gamma^1)$ are conjugate to one another in the bulk theory. This gives us the following braiding relation between the operators.
\beq
\cU_\alpha(M^3)\cV_n(\gamma^1) = e^{2\pi i\alpha n\,\ell(\gamma^1,M^3)}\cV_n(\gamma^1)\cU_\alpha(M^3)
\eeq
Here we are using $\ell(\gamma^1,M^3)$ to denote the linking of the submanifold $\gamma^1$ and $M^3$ within $\cM^5$ \cite{Bott1982}. As we move to the conformal boundary, we see the local operator $\overline{\cV}_n(x)$ inherits these braiding rules with $\cU_\alpha(M^3)$, giving us that it is charged under the $U(1)$ topological defect. Hence, we have found that the local operator holographically dual to a D3-brane wrapped on $\bR_{+}\times N^3$ acts as a charged object under the symmetry generated by an operator dual to the D3-brane wrapping $M^3\times S^1$ defined in the previous subsection.

%% file: sec_general_SE7.tex

\section{General Sasaki-Einstein in M-theory}\label{sec:general_SE7}

The second example we will discuss is the $R$-symmetry present in $\cN=2$ SCFTs constructed from M-theory on an $AdS_4\times SE_7$ background. Similar to the Type IIB case, this $R$-symmetry is universal, coming from the $U(1)$ isometry of the $SE_7$ when viewed as a circle fibration. Our approach is the same as above: we start with a general consistent truncation and further truncate the gauge fields to the minimal set needed to contain the gauge field associated to the isometry. We then construct the $R$-symmetry generator explicitly and identify the wrapped brane that reproduces it in the boundary field theory. 

\subsection{M-theory KK Reduction}

Our starting point is an 11d geometry of the type described in the previous subsection. The general Kaluza-Klein ansatz for the 11-dimensional metric of the form $AdS_4\times SE_7$ is given by \cite{Gauntlett:2009zw,Bah:2010yt}
\beq
ds_{11}^2 = e^{-6U - V}ds^2(\cM^4) + e^{2U}ds^2(KE_6) + e^{2V}(\eta + \sfA_1)^2.
\eeq
The modes $U$ and $V$ again parameterize different breathing and squashing modes of the K\"ahler-Einstein base, and thus we demand $U=V=0$ to focus on the Sasaki-Einstein case.

The M-theory 4-form $G_4$ is the only 11d field strength that needs to be expanded, and most generally can be written
\begin{align}
G_4 &= f_0{\rm vol}(\cM^4) + g_3\wedge(\eta + \sfA_1) + g_2\wedge J + e^Z dZ\wedge J\wedge (\eta + \sfA_1) + 2e^Z J^2\nn\\
&\qquad +\left[X(\eta+\sfA_1)\wedge\Omega - \frac{i}{4}(dX-4iX\sfA_1)\wedge\Omega + {\rm c.c}\right],
\end{align}
where $X$ is a complex scalar field, $Z$ is a real scalar field, and all other fields are real with degree denoted by their subscript. From the equation of motion for $G_4$, 
\begin{equation}
    d\star G_4 + \frac{1}{2}G_4\wedge G_4 = 0,
\end{equation}
one can find the relation $f_0 = 6(\epsilon + e^{2Z} + \frac{1}{3}|X|^2)$ for $\epsilon = \pm 1$. The $\epsilon = +1$ case preserves supersymmetries while the $\epsilon = -1$ solution does not and is known as the ``skew-whiffed'' solution. We will take the positive solution moving forward.

We are concerned with the {\it universal} 4d $\cN=2$ multiplet obtained from this background containing the metric, the $U(1)$ isometry gauge field $\sfA_1$, and their supersymmetric partners. To truncate consistently to this multiplet, we set 
\begin{equation}
    e^Z = X = g_3 = 0,\qquad g_2 = -\star_4\sfF_2.
\end{equation}
This then yields $f_0 = 6$, and reduces $G_4$ to 
\begin{align}\label{eq:Mtheory_RRflux}
G_4 &= 6{\rm vol}(\cM^4) - (\star_4\sfF_2)\wedge J\nn\\
&= 6{\rm vol}(\cM^4) + \star_4\sfF_2\wedge J + \sfF_2\wedge\star_4\sfF_2 - d\left(\star_4\sfF_2\wedge(\eta+\sfA_1)\right).
\end{align}

With our truncation specified, and thus our field content, we are in a position to construct the bulk theory as discussed in Section \ref{sec:branes_SymOps}. Our starting point is the 12-form M-theory anomaly polynomial \cite{Freed:1998tg,Harvey:1998bx,Bah:2019rgq}
\begin{equation}\label{eq:Mtheory_anom_poly}
    \cI_{12} = -\frac{1}{6}G_4\wedge G_4\wedge G_4 - G_4\wedge X_8,
\end{equation}
where $X_8$ is an 8-form constructed from data on the tangent bundle of spacetime as follows.
\begin{equation}
    X_8 = \frac{1}{192}\left(p_1(TM^{11})\wedge p_1(TM^{11}) - 4p_2(TM^{11})\right)
\end{equation}
This 8-form is primarily related to gravitational anomalies coming from the allowed structures on the spacetime, and thus is not immediately relevant to our purposes of studying the $U(1)$ $R$-symmetry. As such, we omit the $X_8$ term from our analysis. By reducing \eqref{eq:Mtheory_anom_poly} over the internal $SE_7$ geometry, we find a 4d bulk theory given by
\begin{equation}\label{eq:Mthy_bulk}
    S_{\rm 4d} = \int_{\cM^4}\frac{1}{6}\sfF_2\wedge\star_4\sfF_2.
\end{equation}
We may now obtain our Page charge either from a timeslice quantization of $S_{\rm 4d}$ and finding the imposed constraints, or from enforcing the M-theory Bianchi identity $dG_4 = 0$. Both methods will yield us 
\begin{equation}
    \cG_{\sfA_1} = d\star_4\sfF_2 = 0 \quad \Rightarrow \quad P_{\sfA_1} = \star_4\sfF_2,
\end{equation}
from which can construct the symmetry operator from the holonomy of $P_{\sfA_1}$ as
\begin{equation}\label{eq:Mtheory_op}
    \cU_\alpha(M^2) = \exp\left(2\pi i\int_{\cM^3}\lambda_0\,\cG_{\sfA_1}\right) = \exp\left(2\pi i\alpha\int_{M^2}P_{\sfA_1}\big|_{\cM^3}\right),
\end{equation}
where $\cM^3$ is the slice upon which we are performing the quantization. The constant $\alpha$ coming from a singular gauge transformation $d\lambda_0 = \alpha\,\delta(M^2)$ can be taken to be valued in $[0,1)$ due to the quantization of the Page charge, and so we have indeed found a topological $U(1)$-valued operator.

\subsection{M2-brane on the $S^1$ Fiber}

With the expression for $\cU_\alpha(M^2)$ obtained in the previous subsection, we now aim to match it with a brane configuration in the bulk M-theory. We again need a brane that preserves the $U(1)$ isometry in the $SE_7$ fibration and is codimension-2 with respect to the bulk $\cM^4$. Due to there being fewer known branes in M-theory when compared to Type IIB, we luckily have fewer configurations to scan through. 

Our probe brane of interest in this set up is an M2-brane wrapping the $S^1$ fiber in the internal $SE_7$. We obtain our desired operator by then reducing \eqref{eq:Mtheory_WZ} over the specified internal directions. Using the expression for $G_4$ derived above in \eqref{eq:Mtheory_RRflux}, we compute the reduction as follows.
\beq
\cI_2 = \int_{S^1}C_3 = \star_4\sfF_2
\eeq
Making a necessary choice of orientation, we find the holonomy of $\cI_2$, and thus the symmetry operator corresponding to the wrapped M2-brane, to be
\beq\label{eq:Mthy_defect}
\cU_\alpha(M^2) = \exp\left(2\pi i\alpha\int_{M^2}\star_4\sfF_2\right),
\eeq
exactly matching the expected operator derived previously. The presence of $\alpha$ should be understood in the same way that we argued at the end of Section \ref{sec:general_SE5}. From this we see that is naturally takes values in $[0,1)$ due to the quantization of the Page charge $P_{\sfA_1}$. 

\subsection{Charged Operator from Wrapped M5-brane}

Now that we have obtained a topological defect \eqref{eq:Mthy_defect}, we should see what is charged under it. As we are saying that $\cU_\alpha(M^2)$ is the symmetry generator for a $U(1)$ 0-form symmetry, the charged objects should be local operators in the boundary QFT. With this in mind, let us consider an M5-brane with worldvolume $\bR_+\times N^5$, where $\bR_+$ denotes the radial direction of $\cM^4$ and $N^5$ is comprised of the $S^1$ fiber as well as the four-dimensional manifold dual to the 4-form $J\wedge J$ in the $SE_7$ internal space.

In the absence of any worldvolume flux for the self-dual 2-form $B_2$, the WZ action \eqref{eq:Mtheory_WZ} for the M5-brane can be reduced on $N^5$ as follows, up to a choice in orientation.\footnote{
Recall that $C_6$ is the gauge field corresponding to the field strength $G_7 = \star G_4$, so that the expression for $C_6$ can be derived directly from \eqref{eq:Mtheory_RRflux}.}
\beq
\cI_1 = \int_{N^5}C_6 = \sfA_1
\eeq
We thus find that the M5-brane describes a line operator in $\cM^4$,
\beq
\cV_1(\gamma^1) = \exp\left(2\pi i \int_{\gamma^1}\sfA_1\right),
\eeq
where $\gamma^1$ is along the radial direction $\bR_+$. As above, we may consider a stack of $n$ M5-branes to recover the operator $\cV_n(\gamma^1)$ for $n\in\bZ$. This line operator, or more accurately its endpoint, seems to be trivial on the conformal boundary $\d\cM^4$ in the case where $\sfA_1$ is equipped with Dirichlet boundary conditions, which is the necessary case if we wish the $R$-symmetry to be a {\it global} symmetry in the boundary theory. However, as in the Type IIB case above, we can write the boundary conditions as
\beq
\left.\sfA_1\right|_{\d\cM^4} = d\varphi_0 = 0,
\eeq
for $\varphi_0$ a flat $U(1)$-valued scalar on the boundary. This allows us to write the endpoint of the M5-brane on the conformal boundary as a non-trivial, non-topological local operator $\overline{\cV}_n(x\in \d\gamma^1)$.

Now that we have constructed a local operator $\overline{\cV}_n(x)$ in the boundary QFT, it remains to be seen if it is charged under $\cU_\alpha(M^2)$ in the appropriate sense. We look at the braiding relations of the extended operators $\cU_\alpha(M^2)$ and $\cV_n(\gamma^1)$ in $\cM^4$, using the commutation relations derived from the bulk theory \eqref{eq:Mthy_bulk}. One can find that the momentum conjugate to the gauge field $\sfA_1$ is simply
\beq
\frac{\Pi_{\sfA_1}}{2\pi} = \star_4\sfF_2,
\eeq
which is exactly the integrand of $\cU_\alpha(M^2)$. This allows us to define a braiding relation between the bulk operators in $\cM^4$ as
\beq
\cU_\alpha(M^2)\cV_n(\gamma^1) = e^{2\pi i \alpha n\,\ell(\gamma^1,M^2)}\cV_n(\gamma^1)\cU_\alpha(M^2).
\eeq
This relation is inherited by $\cU_\alpha(M^2)$ and $\overline{\cV}_n(x)$ on the boundary $\d\cM^4$, thus giving us the desired behavior. Namely, the local operator living at the end of the M5-brane wrapped on $\bR_+\times N^5$ is charged under the symmetry generator described by an M2-brane wrapping $M^2\times S^1$.

As an interesting note, we can use the M2- and M5-brane configurations discussed in this section to reproduce the two D3-brane configurations discussed in Section \ref{sec:general_SE5} as a consistency check. The duality chain is as follows.
\beq
    \begin{gathered}
        \cU_\alpha(M^2) \to {\rm M2}\xrightarrow{\widetilde{S^1}\; {\rm reduction}} {\rm D2} \xrightarrow{\rm T-duality} {\rm D3} \to \cU_\alpha(M^3)\\
        \cV_1(\gamma^1) \to {\rm M5}\xrightarrow{\widetilde{S^1}\; {\rm reduction}} {\rm D4} \xrightarrow{\rm T-duality} {\rm D3'} \to \cV_1(\gamma^1)\\
    \end{gathered}
\eeq
In words, we can first reduce our 11d M-theory on an $\widetilde{S^1}$ submanifold of the $KE_6$ base of our internal $SE_7$, parallel to the M5-brane used to construct $\cV_1(\gamma^1)$. This would result in our M2- and M5-branes respectively describing D2- and D4-branes in a Type IIA background. We then perform a T-duality transformation along a dimension of the internal geometry parallel to the D4-brane as well as an external direction orthogonal to the D2-brane. Our final result in Type IIB is then the two D3-branes we used to construct the $\cU_\alpha(M^3)$ and $\cV_1(\gamma^1)$ operators in Section \ref{sec:general_SE5}, corresponding respectively to the $\cU_\alpha(M^2)$ and $\cV_1(\gamma^1)$ operators we have constructed in the present section.

%% file: sec_discussion.tex

\section{Discussion}\label{sec:discussion}

In this paper we have discussed the topological defects describing the $U(1)$ $R$-symmetry in 4d and 3d SCFTs constructed holographically in Type IIB string theory and M-theory respectively. We first derived the form of these operators using a theory defined in the bulk space on whose boundary the SCFT lives, coming from the respective 11- and 12-form anomaly polynomials of the bulk theory. As the symmetry operators come from Page charges defined in this theory, their topological nature on the boundary is manifest. Moreover, we have constructed the symmetry operators using wrapped probe branes defined in the bulk geometry. By pushing the wrapped branes to the conformal boundary, the worldvolume dynamics decouple and we recover the Page charges from the remaining Wess-Zumino action. We conclude in this section by discussing some interesting relations between our results and other results in the literature, as well as propose some interesting future directions.

Something we did not discuss much in the previous sections is that the exact expressions for $P_{\sfA_1}$ when restricted to $\partial\cM^{q+2}$ are not strictly defined on the boundary in a way independent of the bulk. That is, there is an explicit Hodge star of the bulk theory present in the expression for the Page charge. To be more precise in the above expressions, we should write instead the integrand of \eqref{eq:IIB_operator} and \eqref{eq:Mtheory_op} as a closed $q$-form $\Lambda_q$ defined in the boundary spacetime $\partial\cM^{q+2}$, where $q=\{3,2\}$ in the Type IIB and M-theory constructions respectively. In both cases this $q$-form is dual, in the boundary theory, to a closed 1-form. It it tempting to identify this closed 1-form as the field strength for a singleton mode $\phi$ propagating on the conformal boundary of the bulk spacetime \cite{Gukov:2004id,Maldacena:2001ss,Bah:2020uev}. This pairs nicely with the discussion of symmetry operators from brane stacks as being described by the center-of-mass modes of the branes as discussed in \cite{Bah:2023ymy}. There, the authors identified the symmetry operator realized by a stack of D5-branes as coming from the center-of-mass mode for the stack. This then worked to explain the fusion rules of the associated operators in the field theory. Further study of singletons arising in this way could prove useful for further developing a bulk/boundary dictionary in holographic constructions. It is believed that singleton modes of the bulk theory should account for many modes in the boundary QFT that decouple in the deep IR. A systematic investigation into wrapped brane configurations as we have constructed in the present paper could provide a way to study the singleton modes in a holographic theory.

Another curious direction is how similar approaches could be used to account for gravitational symmetries and their anomalies in holographic QFTs. In 3d theories for example, like those obtained from an M-theory $AdS_4\times SE_7$ background, there is the possibility of a parity anomaly coming from Chern-Simons contact terms \cite{Closset:2012vp}. The authors of \cite{vanBeest:2022fss} studied a specific example wherein the cancellation of the parity anomaly affected the global higher-form symmetries of the theory. It would be interesting to see if the so-far-neglected contributions from $X_8$ could capture this information, and, in the cases where the QFT is parity symmetric, if the associated symmetry defects could be realized by wrapped M-branes.

The consistent truncations discussed in the present paper are only the minimal ones, in the sense that we have truncated to only the universal gravity multiplet in each case. Of course, one may consider more involved consistent truncations wherein more fields are present in the reduction. In such constructions one may find that previously-massless gauge fields become massive vector fields due to, for example, a St\"uckelberg mechanism. This can result in wrapped branes probing global symmetries that are {\it broken} in the dual field theory \cite{Cvetic:2025kdn}. However, there still may be a massless gauge field yielding a global symmetry generator, arising as a linear combinations of the vector fields present in the truncation. A general analysis of the role of massive gauge fields in consistent truncations could lead to a more complete understanding of broken and preserved symmetries in holographic constructions.

Finally, we have only discussed cases where the global symmetry in the boundary theory is given by a single $U(1)$ factor. It would be very useful to extend the present discussion to theories with continuous non-Abelian symmetries, or further to theories with continuous higher structures such as 2-groups (or more general $n$-groups) or a continuous analogue of higher fusion categories.